\documentclass[epj]{svjour}
\usepackage{graphicx}
\usepackage{placeins}
\usepackage{hhline}
\usepackage{xcolor}
\usepackage{amssymb, amsmath}
\newcommand{\rns}{\rho_{\rm sat}}
\newcommand{\Ms}{M_{\odot}}

\newcommand{\Mc}{\mathcal{M}_c}

\newcommand{\procspie}{Proceedings of the SPIE} 
\newcommand{\apj}{ApJ} 
\newcommand{\apjl}{Astrophys. J. Letters} 
\newcommand{\nat}{Nature} 
\newcommand{\physrep}{Physics Reports} 
\newcommand{\prd}{Phys. Rev. D}
\newcommand{\prc}{Phys. Rev. C}
\newcommand{\aap}{Astronomy \& Astrophysics} 
\newcommand{\mnras}{Monthly Notices of the Royal Astronomical Society}

\begin{document}

\title{Constraints on the Neutron Star Equation of State from GW170817}
\titlerunning{Constraints on the Neutron Star EOS from GW170817}
\author{Carolyn A. Raithel}
\institute{Department of Astronomy and Steward Observatory, University of Arizona, 933 N. Cherry Avenue, Tucson, Arizona 85721, USA}
\date{Received: date / Revised version: date}

\abstract{
The first detection of gravitational waves from a neutron star-neutron star merger, GW170817, has opened up a new avenue for constraining the ultradense-matter equation of state (EOS). The deviation of the observed waveform from a point-particle waveform is a sensitive probe of the EOS controlling the merging neutron stars' structure. In this topical review, I discuss the various constraints that have been made on the EOS in the year following the discovery of GW170817. In particular, I review the surprising relationship that has emerged between the effective tidal deformability of the binary system and the neutron star radius. I also report new results that make use of this relationship, finding that the radius inferred from GW170817 lies between 9.8 and 13.2~km at 90\% confidence, with distinct likelihood peaks at 10.8 and 12.3~km. I compare these radii, as well as those inferred in the literature, to X-ray measurements of the neutron star radius. I also summarize the various maximum mass constraints, which point towards a maximum mass $\lesssim2.3~\Ms$, depending on the fate of the remnant, and which can be used to additionally constrain the high-density EOS. I review the constraints on the EOS that have been performed directly, through Bayesian inference schemes. Finally, I comment on the importance of disentangling thermal effects in future EOS constraints from neutron star mergers.
\PACS{
      {97.60.Jd}{Neutron stars}   \and
      {26.60.Kp}{equations of state} \and
      {95.85.Sz}{astronomical observations of gravitational waves}
     } 
} 

\maketitle

\section{Introduction}

On August 17, 2017, the first detection of gravitational waves from a neutron star-neutron star merger heralded the start of a new era of astrophysics. The event, GW170817, was observed by the two Advanced LIGO and one Advanced Virgo detectors, with a matched-filter signal-to-noise ratio of 32.4. The signal remained in the detectors' sensitive band for $\sim100$~s, corresponding to $\sim3000$ orbital cycles in the initial frequency band considered \cite{Abbott2017a}. Following the binary coalescence  by 1.7~s, the gamma-ray burst (GRB), GRB 170817A, was observed independently by the Fermi Gamma-ray Burst Monitor and the International Gamma-Ray Astrophysics Laboratory, in the same direction as GW170817 \cite{Abbott2017c}. Roughly 11 hours later, a fading optical transient was discovered by the Swope Supernova Survey (SSS17a/AT2017gfo) \cite{Coulter2017}, which was extensively followed up \cite{Abbott2017b}. Early observations indicated a blue transient that faded over the initial 48~hours, followed by significant reddening over the next $\sim10$~days. Altogether, these observations indicate that GW170817 was the gravitational signature produced by a neutron star-neutron star merger, which subsequently produced a short-GRB and a kilonova, the latter powered by the radioactive decay of matter in the ejecta (see Ref.~\cite{Abbott2017b} and references therein).

The initial analysis by the LIGO-Virgo collaboration used a matched-filter analysis to determine the parameters of a tidally-corrected post-Newtonian waveform model. The component masses of GW170817 were determined to be in the range $1.17-1.60~\Ms$, with a total binary mass of $2.74\substack{+0.04\\-0.01}~\Ms$. The chirp mass, defined as
\begin{equation}
\Mc = \frac{(m_1 m_2)^{3/5}}{(m_1+m_2)^{1/5}},
\end{equation}
where $m_1$ and $m_2$ are the component masses, was very tightly constrained, to $\Mc=1.188\substack{+0.004\\-0.002}~\Ms$, for the case of low-spin priors \cite{Abbott2018}.\footnote{Throughout this paper, I will only report results for the case of low-spin priors, as these are most consistent with the spin distribution inferred for galactic binary neutron stars.}

Of particular interest to the neutron star community, was the accompanying first measurement of equation-of-state (EOS) effects on the gravitational wave signal. The waveform produced by the coalescence  of two neutron stars should, in principle, deviate from a point-particle waveform, due to tidal effects on the neutron star matter. In one early study, Ref.~\cite{Read2009a} found that these differences could potentially be measured with Advanced LIGO, and that the degree of deviation could be used to differentiate between EOS that differ in the radii they predict for neutron stars by only $\sim1$~km \cite{Read2009a,Read2013,Lackey2015}.

In anticipation of future gravitational wave detections, Ref.~\cite{Flanagan2008} showed that the EOS-dependent effects during the early phase of the inspiral could be cleanly represented by a single parameter:  the tidal Love number, $\lambda$. The tidal Love number is defined as the ratio of the tidally-induced quadrupole deformation, $Q^{(\rm tid)}$, to the tidal potential caused by the binary companion, $\epsilon^{(\rm tid)}$, i.e., 
\begin{equation}
\lambda \equiv -\frac{Q^{(\rm tid)}}{\epsilon^{(\rm tid)}},
\end{equation}
or, in its dimensionless form, 
\begin{equation}
\Lambda \equiv \frac{\lambda}{M^5} \equiv \frac{2}{3}k_2^{(\rm tid)} C^{-5},
\label{eq:Lambda}
\end{equation}
where $M$ is the stellar mass and $C\equiv GM/Rc^2$ is the stellar compactness, with $R$ as the radius and $G$ and $c$ the gravitational constant and speed of light, respectively. Here, I follow the convention of Ref. \cite{Flanagan2008} and call $k_2^{\rm(tid)}$ the tidal apsidal constant. The tidal apsidal constant depends both on the stellar compactness and the particular EOS, and has been constrained to $0.05\lesssim k_2^{(\rm tid)} \lesssim 0.15$ for a variety of realistic, hadronic EOS \cite{Hinderer2008,Hinderer2010,Postnikov2010}. Thus, a measurement of $\Lambda$ offers possible insight to the underlying EOS.

The phase of a gravitational-wave signal is determined, to first order, by the related parameter $\widetilde{\Lambda}$, which is the effective tidal deformability of the binary system. The effective tidal deformability is defined as
\begin{equation} 
\widetilde{\Lambda} \equiv \frac{16}{13} \frac{ (m_1 +  12 m_2) m_1^{4} \Lambda_1 + (m_2 + 12 m_1) m_2^{4} \Lambda_2}{(m_1+m_2)^5}, 
\label{eq:Leff}
\end{equation}
where $\Lambda_1$ and $\Lambda_2$ are the dimensionless tidal deformabilities of the component stars \cite{Flanagan2008,Favata2014}.

The initial analysis of GW170817 by the LIGO-Virgo Collaboration inferred $\widetilde{\Lambda}\le900$ at the 90\% confidence level,\footnote{The initial analysis in Ref.~\cite{Abbott2017a} quoted a 90\% upper limit of $\widetilde{\Lambda}\le 800$; however, this was an error. The correct upper limit for the original TaylorF2 waveform and low-spin priors is $\widetilde{\Lambda}\le 900$, as discussed in the caption of Table~IV in Ref.~\cite{Abbott2019}. All of the analyses discussed in this review used the initial upper limit of 800 where relevant. } and showed that this disfavors stiff EOS that predict less compact (i.e., more extended) stars  \cite{Abbott2017a,Abbott2019}. This measurement prompted a flurry of papers, investigating the implications for the neutron star EOS. Roughly one year after the initial detection, the LIGO-Virgo Collaboration released a re-analysis of the data from GW170817. The re-analysis provided in Ref.~\cite{Abbott2019} improved on the initial constraints by re-calibrating the Virgo data, extending the range of frequencies included, using a new set of waveform models that go beyond the post-Newtonian approximation, and incorporating source distance measurements from the electromagnetic counterpart. With the new analysis, the chirp mass was revised down slightly to $\Mc=1.186\substack{+0.001\\-0.001}~\Ms$ and more detailed information on the tidal deformability was provided. In particular, they reported a 90\% highest posterior density interval of $\widetilde{\Lambda} = 300\substack{+420\\-230}$. 

In this paper, I will review the status of EOS constraints that have been published since the discovery of GW170817 and its initial analysis. I will focus on EOS constraints inferred primarily from the gravitational wave signal, although I note that there is rich information to be learned also from study of the electromagnetic counterpart (see, e.g., Refs.~\cite{Chornock2017,Kasen2017,Cote2018,Hotokezaka2018} for the implications of GW170817 for r-process nucleosynthesis) as well as from a broader exploration of the gravitational wave data (see, e.g., Refs.~\cite{Malik2018,Zhang2018,Carson2019,Zhang2019} for constraints on the nuclear symmetry energy or Refs.~\cite{Montana2018,Paschalidis2018,Alvarez-Castillo2019,Christian2019} for discussion of GW170817 and so-called ``twin stars").

I will start with a brief overview of EOS models and existing constraints. In $\S$\ref{sec:tidal}, I will review a series of papers that have found a direct link between $\widetilde\Lambda$ and the neutron star radius. In $\S$\ref{sec:Mmax}, I will review new constraints on the upper limit of the maximum mass of neutron stars, and their implication for the EOS. In $\S$\ref{sec:Bayes}, I will discuss findings from Bayesian inference schemes that infer EOS parameters directly from the gravitational wave data.  Finally, I will comment on the importance of disentangling thermal effects from the cold EOS in future work.

\section{Existing equation of state constraints}
\label{sec:EOS}
The EOS of cold, ultra-dense matter remains poorly constrained at high densities, despite decades of study. A wide variety of theoretical models have been proposed, ranging from pure nucleonic models (e.g., \cite{Baym1971,Friedman1981,Akmal1998,Douchin2001}) to models predicting more exotic phases of matter, such as pion condensates (e.g., \cite{Pandharipande1975}), kaon condensates (e.g., \cite{Kaplan1986}), hyperons (e.g., \cite{Balberg1997}), or deconfined quark matter at high densities (e.g., \cite{Collins1975}). More recently, some studies have started to incorporate these quark degrees of freedom using state-of-the-art results from perturbative QCD (e.g., \cite{Fraga2014}).  However, despite the diversity of these models, it remains possible that they do not cover the full range of allowed physics. This possibility has motivated the creation of a large number of parametric EOS models. Among these models are the piecewise polytropic or linear representations \cite{Read2009,Ozel2009,Hebeler2010,Steiner2010,Raithel2016}, spectral expansion methods \cite{Lindblom2010}, and constant speed-of-sound parameterizations \cite{Alford2013}.  

Nuclear experiments are only able to constrain these various models up to densities near the nuclear saturation density, $\rns = 2.7\times10^{14}$~g/cm$^3$;  extrapolations to higher densities remain difficult \cite{Lattimer2012}. Moreover, while the interactions between particles can be written in terms of few-body potentials up to $\sim\rns$ \cite{Akmal1998,Morales2002,Gandolfi2012}, the expansion of interactions into few-body terms starts to break down at higher densities, due in part to the overlap of the particle wave-functions \cite{Ozel2016}.

Observations of neutron stars offer one way to constrain the ultra-dense EOS, with densities in the cores of neutron stars reaching $8-10\rns$. Macroscopic observations of neutron star masses \cite{Demorest2010,Antoniadis2013,Fonseca2016} and radii \cite{Ozel2009a,Guver2010,Guillot2013,Guillot2014,Heinke2014,Nattila2016,Ozel2016a,Bogdanov2016,Ozel2016}  can be used to constrain realistic or parametric EOS, by taking advantage of the one-to-one mapping between the EOS and the mass-radius relation \cite{Lindblom1992}. Reference~\cite{Ozel2010} performed the first inference of the EOS from X-ray observations. They used a sample of three radii measured from neutron stars undergoing thermonuclear bursts to infer the parameters of piecewise polytropic EOS. Reference~\cite{Steiner2010} performed a similar inference for a sample extended to include three additional radii measured from transient low-mass X-ray binaries. In Ref.~\cite{Steiner2013}, the sample was again extended to eight sources and a wider variety of parametric EOS were studied, including piecewise polytropic and piecewise linear models, as well as models allowing quark matter at high densities. For their baseline polytropic model, the authors constrained the radius of a $1.4\Ms$ star to the range $10.4-12.9$~km. The inferred EOS constraints are recreated in Fig.~\ref{fig:EOS}.

New measurements over the past few years have further increased the sample of X-ray radii. In a recent analysis, Ref.~\cite{Ozel2016} used a Bayesian inference scheme to combine fourteen radius measurements from X-ray bursts and quiescent low-mass X-ray binaries,  while simultaneously requiring consistency with low-energy nucleon-nucleon scattering data as well as a maximum mass that is consistent with the observations of the most massive neutron stars \cite{Demorest2010,Antoniadis2013,Fonseca2016}. They inferred the parameters of a three-polytrope EOS model, which I recreate in Fig.~\ref{fig:EOS}, and constrained $R(1.5~\Ms)$ to lie in the range $9.9-11.2$~km \cite{Ozel2016}, slightly below the range inferred by Ref.~\cite{Steiner2013}. With a more restrictive assumption of a mono-parametric EOS, i.e., assuming that all neutron stars have the same radius, Ref.~\cite{Ozel2016a} use a sample of 12 radius measurements to infer that the neutron star radius is $10.3\pm0.5$~km. 

While these radii measurements are starting to converge, there remain uncertainties. For a recent review of relevant systematic uncertainties and open questions, see Ref.~\cite{Ozel2016}. The in-progress NASA Neutron Star Interior ExploreR (NICER) mission \cite{Gendreau2012}  will provide radii measurements from pulse profile modeling of X-ray pulsars, with uncertainties potentially as small as 0.5~km. These anticipated data may provide more stringent constraints on the EOS. Additionally, a measurement of the moment of inertia from the double pulsar system, J0737$-$3039 \cite{Lyne2004}, is expected at 10\% accuracy in the coming years and will offer independent constraints on the EOS \cite{Morrison2004,Lattimer2005,Kramer2009,Raithel2016a,Kehl2016}.

The advent of gravitational wave observations offers a complementary avenue for constraining the EOS, independent of any of the systematics which may have biased existing measurements. Moreover, the tidal deformability represents a measurement of the quadrupolar mass distribution of the star, which has never before been measured and thus offers fresh insight into the neutron star interior. I will now turn to these constraints and insights for the remainder of the paper.

\section{Tidal deformability as a probe of the neutron star radius}
\label{sec:tidal}
The tidal deformability measures the tidally-induced deformation of a neutron star, relative to the tidal potential of its binary companion. As such, it contains rich information about the stellar structure. Equation~(\ref{eq:Lambda}) shows that the dimensionless tidal deformability, $\Lambda$, can be written simply in terms of the stellar compactness and the tidal apsidal constant, which in turn depends on the compactness and the particular EOS. However, the tidal apsidal constant has been constrained to a relatively narrow range, $ 0.05 \lesssim k_2^{\rm(tid)} \lesssim 0.15$, for realistic, hadronic EOS \cite{Hinderer2008,Hinderer2010,Postnikov2010}.

In a pivotal study, Ref. \cite{Yagi2013} showed that the narrow range of $k_2^{(\rm tid)}$ corresponds more generally to a quasi-universal relation between $\Lambda$ and the stellar compactness that holds for a variety of EOS. This relation is a corollary of the so-called ``I-Love-Q" universal relations between the moment of inertia, tidal Love number, and quadrupole moment of neutron stars. Subsequently, Ref.  \cite{Yagi2017} found that the $\Lambda-C$ relationship can be written, for an even larger sample of EOS, as
\begin{equation}
\label{eq:YY}
C = a_0 + a_1 \ln\Lambda + a_2 (\ln\Lambda)^2,
\end{equation}
where the coefficients were fit to be $a_0=0.360, a_1=-0.0355$, and $a_2=0.000705$, with errors of $\lesssim$6.5\%.

Reference \cite{Annala2018} extended the work to a sample of 260,000 EOS, which they constructed to match chiral effective theory results at low densities and perturbative QCD results at high densities. In between these two regimes, the EOS were allowed to span the entire thermodynamically-consistent phase space, with either 3 or 4 polytropic segments. They found that, for a star with a fixed mass of a $1.4~\Ms$,
\begin{equation}
\Lambda(1.4\Ms) = 2.88 \times 10^{-6} \left(\frac{R(1.4 \Ms)}{\rm km}\right)^{7.5},
\end{equation}
essentially extending the universal relation in compactness to apply more directly to radius and for a broader range of EOS.  Combined with the initial gravitational wave constraint on $\Lambda(1.4~\Ms) < 800$ from Ref. \cite{Abbott2017a}, the authors constrained the radius of a $1.4~\Ms$ star to be less than 13.6~km, marking one of the first constraints on the neutron star radius from GW170817.

Similar analyses have been performed based on state-of-the-art crustal models at low densities and perturbative QCD at high densities \cite{Most2018} or chiral effective field theory models \cite{Lim2018,Tews2018}, all of which calculated a large array of EOS models, subject to these various nuclear constraints. The ranges of tidal deformabilities predicted by the constrained EOS models were found to be consistent with the tidal deformability inferred from GW170817, but these studies point to the potential constraining power of future detections. In particular, Refs.~\cite{Most2018,Lim2018} highlighted that a lower limit on $\widetilde{\Lambda}$ would provide strong constraints on the EOS.

In parallel to these early estimates of $R$ from the component tidal deformabilities of GW170817, it was discovered that the effective tidal deformability of the  \textit{binary} system can, surprisingly, also be used as a direct probe of the radius. The mass dependence that enters $\widetilde{\Lambda}$ (eq.~\ref{eq:Leff}) both explicitly through the masses of the component stars, as well as implicitly through $\Lambda_1$ and $\Lambda_2$, was found to be extremely weak, when the chirp mass is fixed. Because $\Mc$ is tightly constrained for GW170817, this feature renders $\widetilde{\Lambda}$ a direct probe of the radius.

Early hints of the potential of using $\widetilde{\Lambda}$ and $\Mc$ to constrain the radius can be found in Ref.~\cite{Wade2014}. They showed that a dimensionful radius-like parameter can be constructed from $\widetilde{\Lambda}$, so that a measurement of $\Mc$ and $\widetilde{\Lambda}$ can be viewed as a measurement of a chirp mass and ``chirp radius." With this perspective, the inversion of $\Mc-R_c$ to EOS parameters becomes analogous to the traditional inverse stellar structure problem, which relates mass-radius measurements to the EOS \cite{Lindblom1992}.

\begin{figure}[ht]
\centering
\includegraphics[width=0.5\textwidth]{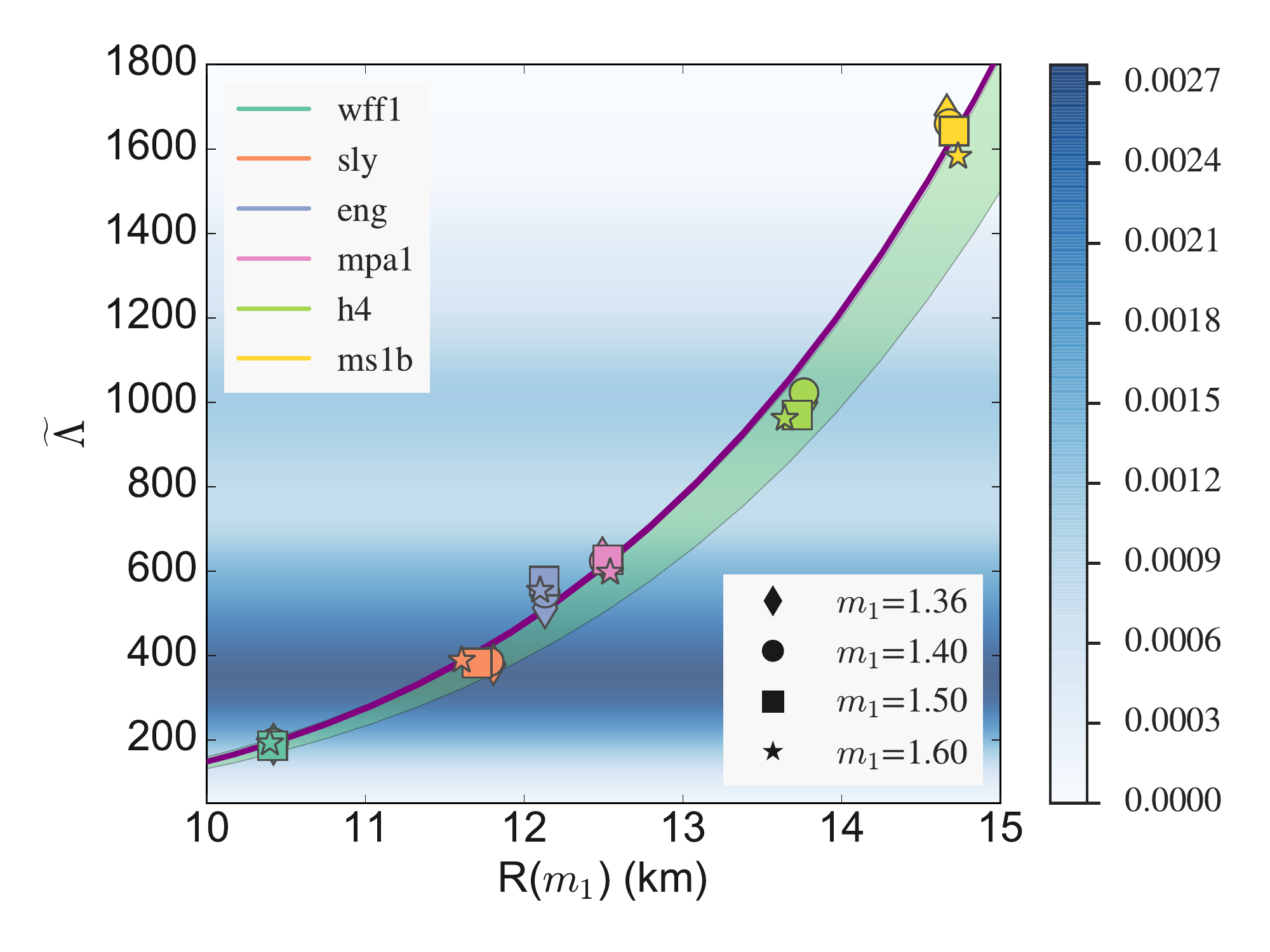}
\caption{\label{fig:LvsR} Effective tidal deformability of the binary system as a function of the radius of the primary neutron star in the merger. $\widetilde{\Lambda}$ is calculated for various primary masses (corresponding to the different symbols) and various EOS (corresponding to the different colors). This figure is similar to Fig.~1 of Ref.~\cite{Raithel2018}, but now uses the revised chirp mass of GW170817 from Ref.~\cite{Abbott2019}, $\Mc=1.186~\Ms$, to calculate the masses of the secondary star. Additionally, overlaid in blue is the one-dimensional posterior distribution of $\widetilde{\Lambda} $ for GW170817, as digitized from Ref.~\cite{Abbott2019} for the PhenomPNRT model. The thin purple swatch corresponds to the quasi-Newtonian expansion of eq.~(\ref{eq:expansion}) for $q=0.7-1.0$, while the light green band corresponds to the approximate relationship of eq.~(\ref{eq:De})~\cite{De2018}.}
\end{figure}

Reference \cite{Raithel2018} showed the first evidence of the surprisingly tight correlation between $\widetilde{\Lambda}$ and the primary neutron star radius, which I recreate in Fig.~\ref{fig:LvsR}. The authors found that, for a wide range of realistic EOS and a fixed chirp mass, the binary tidal deformability correlates strongly with the radius of the primary neutron star, regardless of the individual component masses. This trend can be seen in Fig.~\ref{fig:LvsR}. In this figure, I fix the chirp mass to the central value for GW170817, $\Mc=1.186~\Ms$ \cite{Abbott2019}. I also show, in blue, the one-dimensional posteriors of $\widetilde{\Lambda}$ digitized from the LIGO-Virgo re-analysis in Ref.~\cite{Abbott2019}. Clearly, the tidal deformability from GW170817 points to a small radius of $\lesssim13.5$~km.

Reference \cite{Raithel2018} also explored the origin of this relationship in the quasi-Newtonian regime. In particular, the authors applied a metric correction to the Newtonian tidal deformability of an $n=1$ polytropic EOS to define a ``quasi-Newtonian" version of the effective tidal deformability of the binary system, $\widetilde{\Lambda}_{qN}$. Using a series expansion on $\widetilde{\Lambda}_{qN}$ in terms of the mass ratio, $q\equiv m_2/m_1$, they revealed a very weak dependence of $\widetilde{\Lambda}_{qN}$ on $q$, i.e.,
\begin{equation}
\label{eq:expansion}
\widetilde{\Lambda}_{qN} = \widetilde{\Lambda}_{0} \left( 1 + \delta_{0} (1-q)^2\right) + \mathcal{O}\left((1-q)^3\right),
\end{equation}
where
\begin{equation}
\label{eq:coef}
\widetilde{\Lambda}_{0}  = \frac{15-\pi^2}{3 \pi^2} \xi^{-5} (1-2 \xi)^{5/2},
\end{equation}
\begin{equation}
\label{eq:correction}
\delta_{0} =  \frac{3}{104}(1-2 \xi )^{-2}\left(-10 + 94 \xi  - 83 \xi^2 \right),
\end{equation}
and they introduced 
\begin{equation}
\label{eq:xi}
\xi = \frac{2^{1/5} G \Mc}{ R c^2}
\end{equation}
as an ``effective compactness" \cite{Raithel2018}. The expression for $\widetilde{\Lambda}_{qN}$ is shown as the thin purple band in Fig.~\ref{fig:LvsR} for $q=0.7-1.0$, with $\Mc=1.186~\Ms$.

Equation~\ref{eq:expansion} reveals that the individual component masses enter only at $\mathcal{O}((1-q)^2)$, while $\widetilde{\Lambda}$ depends strongly on the radius and chirp mass. Thus, for a well-constrained chirp mass, eq.~(\ref{eq:expansion}) can be used to solve directly for the radius, with errors of at most $\sim4\%$ for realistic neutron stars \cite{Raithel2018}.

\begin{figure}[ht]
\centering
\includegraphics[width=0.45\textwidth]{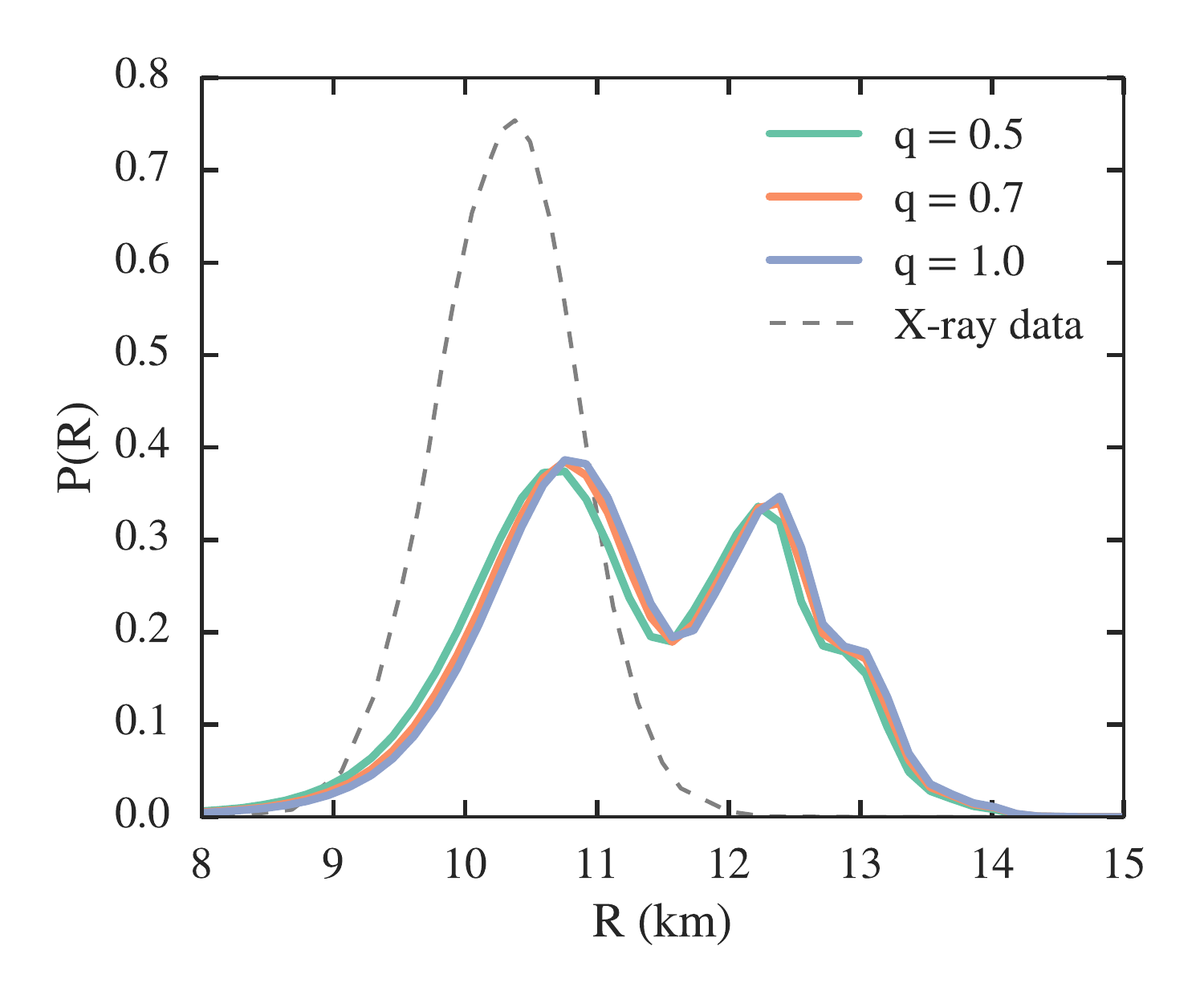}
\caption{\label{fig:Rpdf} Posterior distribution in radius from GW170817 for three different mass ratios, shown in the different colors. I convert the reported posterior distribution for $\widetilde{\Lambda}$ from Ref.~\cite{Abbott2019} to posteriors in radius using the quasi-Newtonian expansion of eq.~(\ref{eq:expansion}). The resulting 90\% highest-posterior density interval corresponds to $9.8 < R < 13.2$~km, for $q=1$, with likelihood peaks at $R\sim$10.8 and 12.3~km. The gray dashed line represents the composite posterior in radius from 12 spectroscopic X-ray measurements, under the assumption that all neutron stars share a common radius \cite{Ozel2016a}. I find approximate agreement between the X-ray data and the higher probability  peak of the inferred radius from GW170817. } 
\end{figure}

One can further use the analytic form of eq.~(\ref{eq:expansion}) to convert the measured posterior distribution in $\widetilde{\Lambda}$ to a distribution in radius, according to
\begin{equation}
P(R) = P(\widetilde{\Lambda}) \biggr\rvert \frac{\partial \widetilde{\Lambda}}{\partial R} \biggr\lvert.
\end{equation}
I show the corresponding distribution in Fig.~\ref{fig:Rpdf}, for a range of mass ratios. I find that the latest posteriors for $\widetilde{\Lambda}$ from GW170817 \cite{Abbott2018} imply a 90\% highest posterior density interval of $9.8 < R < 13.2$~km (for $q=1$), with distinct likelihood peaks at $\sim$10.8~km and 12.3~km. 

This range is consistent with, though slightly broader than that inferred from X-ray radii measurements for polytropic EOS (see $\S$\ref{sec:EOS}). Figure~\ref{fig:Rpdf} also shows, as the gray, dashed line, the radius inferred from a sample of 12 X-ray radii measurements, under the more restrictive assumption that all neutron stars share a common radius with a mono-parametric EOS \cite{Ozel2016a}. This distribution roughly agrees with the higher-significance, smaller-radius peak from GW170187. Finally, Fig.~\ref{fig:Rpdf} indicates that the inferred radii are approximately constant across a wide range of mass ratios, as expected from the weak $q$-dependence discovered in eq.~(\ref{eq:expansion}).

The weak-dependence on the mass ratio was also reported early on in Refs. \cite{De2018} and \cite{Zhao2018}, using a simpler approximation that employs the scaling relation, $\widetilde{\Lambda}\sim C^{-6}$. The authors of both works used the fact that $k_2^{(\rm tid)}$ scales roughly inversely with compactness for masses $\ge1~\Ms$ \cite{Hinderer2010,Postnikov2010},  to show that the dimensionless tidal deformability can be written as
\begin{equation}
\Lambda \simeq a C^{-6}
\end{equation}
where $a \in [0.0086,0.01]$ for a large sample of polytropic EOS \cite{De2018}. The effective tidal deformability of the binary system is then
\begin{equation}
\widetilde{\Lambda} \simeq \frac{16a}{13} \left( \frac{\hat{R} c^2}{G \Mc}\right)^6 f(q),
\end{equation}
where $\hat{R}$ is an effective binary radius and $f(q)$ is a weakly-dependent function of q, defined as
\begin{equation}
f(q) = q^{8/5}(12-11q+12q^2)(1+q)^{-26/5}.
\end{equation}
Neglecting the $q$-dependence of $f(q)$, they then approximate
\begin{equation}
\label{eq:De}
\widetilde{\Lambda} = a\prime \left( \frac{\hat{R}c^2}{G\Mc}\right)^6,
\end{equation}
with $a\prime=0.0042\pm0.0004$. This allows a direct inversion to the radius, according to
\begin{equation}
\label{eq:DeR}
\hat{R} \simeq R_{1.4} \simeq (11.2\pm0.2) \frac{\Mc}{\Ms} \left( \frac{\widetilde{\Lambda}}{800}\right)^{1/6} \rm km
\end{equation}
\cite{De2018}, with similar results found in Ref. \cite{Zhao2018}. I show the relationship predicted by eq.~(\ref{eq:De}) in Fig.~(\ref{fig:LvsR}) as the light green band. This approximate relationship spans a broader range, but is in qualitative agreement with the more rigorous quasi-Newtonian result of Ref. \cite{Raithel2018}.
 
In order to obtain the most robust constraints possible from their $\widetilde{\Lambda}-R$ relationship,
Ref.~\cite{De2018} re-analyzed the data from GW170817, using Bayesian parameter estimation with the added requirement that both stars obey the same EOS, which was not imposed in the initial LIGO-Virgo analysis \cite{Abbott2017a}. They also allowed for different priors on the mass distribution. When using the double neutron star mass distribution as the prior, they found $\widetilde{\Lambda} = 245 \substack{+453 \\-151}$, corresponding, via eq.~(\ref{eq:DeR}), to a common radius of $\hat{R}=10.9 \substack{+2.1\\-1.6}\pm 0.2$~km. Here and in the following estimates, the 0.2~km ``systematic" error comes from the uncertainty in $a\prime$ in the quasi-universal relation of eq.~(\ref{eq:De}).

In an effort to make even tighter, but much more model-dependent constraints, Ref. \cite{Radice2018a} performed a similar parameter estimation on GW170817 and additionally included electromagnetic data. To calculate the likelihood of the electromagnetic observations, the authors used a fitting formula from numerical relativity simulations that relates $\widetilde{\Lambda}$ to the remnant disk mass. Extant light curve models provide an estimate of the observed disk mass, which can then be compared to the value predicted by the simulated formula for a given $\widetilde{\Lambda}$. This method yielded a 90\% confidence interval for $\widetilde{\Lambda}$ of (323,776) with median  487. Using eq.~(\ref{eq:DeR}), the authors converted this to a common radius of $12.2\substack{+1.0\\-0.8}\pm0.2$~km.

In a similar analysis, Ref. \cite{Coughlin2018} performed a parameter estimation using gravitational wave data, combined with radiative transfer models of the kilonova lightcurve, phenomenological fits from numerical relativity simulations of the disk and dynamical ejecta masses, and fits between the GRB jet energy and disk mass. They found $\widetilde{\Lambda} \in [292, 822]$ at the 90\% confidence level, with a common radius inferred via eq.~(\ref{eq:DeR}) of $R \in [11.1, 13.4]$~km, also at the 90\% confidence level, with an additional 0.2~km systematic uncertainty. While these results are in rough agreement with those of Refs. \cite{Raithel2018,De2018,Zhao2018,Radice2018a}, they rely on significant assumptions about the electromagnetic model fitting.

The $\widetilde{\Lambda}-R$ relationship was initially only tested with relatively simple, hadronic EOS. However, it remains possible that neutron-star matter undergoes a phase transition to deconfined quark matter at high densities. Depending on the exact density of the transition, a merger could could contain two neutron stars, two hybrid stars (with nuclear matter at low densities and quark matter in their cores), or one star of each. The latter case, in particular, poses a potential complication for the $\widetilde{\Lambda}-R$ relationship.

Reference \cite{Han2018} explored the effect of such a first-order phase transition on the tidal deformability, using hybrid EOS constructed with a constant-sound-speed parametrization. The authors found a much stronger dependence of $\widetilde{\Lambda}$ on $q$, in the presence of a strong phase transition. In fact, if such a phase transition indeed were present in the EOS, Ref.~\cite{Han2018} showed that it may be possible to observe different values for $\widetilde{\Lambda}$ for mergers with identical chirp masses but different mass ratios. Reference \cite{Burgio2018} similarly showed that the monotonic relationship between $R(1.5~\Ms$) and $\widetilde{\Lambda}$ is violated if there is a strong phase transition in the EOS. This is an important caveat to consider when interpreting the radii presented in this section.

\section{ Constraints on the maximum mass }
\label{sec:Mmax}
The maximum mass of a non-rotating neutron star, $M_{\rm max}$, offers additional constraints on the EOS. Whereas the radii discussed in $\S$\ref{sec:tidal} are set by the pressure near $1-2\rns$ \cite{Lattimer2001}, the maximum mass is determined by the pressure at high densities (e.g., near $\sim8\rns$ for certain EOS \cite{Ozel2009}) rendering it a particularly useful probe of pressures in the high-density regime.

While there exist restrictive lower limits on $M_{\rm max}$ from the observations of two nearly $2~\Ms$ neutron stars \cite{Demorest2010,Antoniadis2013,Fonseca2016}, the upper bounds on $M_{\rm max}$ are much less constraining. A theoretical upper limit can be calculated by assuming a known EOS at low densities and forcing the sound speed to be causal at high densities, as was first demonstrated in Ref. \cite{Rhoades1974}. Setting the transition density at $1.7\rns$ results in $M_{\rm max}\sim3.2~\Ms$ \cite{Rhoades1974}.  With more modern EOS employed up to twice the nuclear saturation density, $M_{\rm max}\sim2.9~\Ms$ \cite{Kalogera1996}. Most modern EOS do not exceed these limits.

There have been attempts to set observational upper limits on $M_{\rm max}$, e.g., from analyses of short GRB formation \cite{Fryer2015,Lawrence2015} or from the mass distribution of binary neutron stars \cite{Antoniadis2016,Alsing2018}. However, these methods are relatively indirect. Observations of GW170817 and its electromagnetic counterpart offer new, independent constraints. These new upper bounds point towards $M_{\rm max} \lesssim 2.3~\Ms$, if the remnant collapses to a black hole early on. In the case of a long-lived neutron star remnant (lifetime $\gtrsim$ days), the maximum mass may be much larger.

Following a merger, the remnant mass, $M_{\rm rem}$, will determine the object's ultimate fate. There are three possible outcomes. If $M_{\rm rem}$ exceeds the maximum mass of a differentially rotating star, the remnant will undergo prompt collapse to a black hole. If $M_{\rm rem} > M_{\rm max}$, but is below the rotating maximum mass, the remnant will survive for a short time as a hypermassive neutron star (HMNS) supported by differential rotation, or as a supramassive neutron star (SMNS) supported by rigid-body rotation. The massive neutron star will spin down until it can no longer be supported through rotational energy, and the object will collapse to a black hole. Finally, if $M_{\rm rem} < M_{\rm max}$, the remnant neutron star may survive indefinitely.

Out of these options, Ref.~\cite{Margalit2017} argued that the remnant must have been a short-lived HMNS, which collapsed to a black hole after $\gtrsim10$~ms. A longer-lived remnant would have transferred significantly more kinetic energy to its surroundings as it spun down, which is inconsistent with the observed energies of the kilonova and GRB afterglow. By combining the total binary mass, inferred from the gravitational wave signal of GW170817, with these energetic constraints, the authors inferred $M_{\rm max} \lesssim 2.17~\Ms$, at the 90\% confidence level.

In contrast, Ref.~\cite{Rezzolla2018} argued for a longer-lived remnant, that is born as a HMNS but collapses to a black hole as a SMNS after redistributing its angular momentum. The authors argued that, in order to produce the observed blue kilonova, the ejecta must have had a large electron fraction, which could only be produced by the hot, polar region of a relatively long-lived remnant. However, the quick production of the short gamma-ray burst ($\sim 1$~s post-merger), implies that the remnant did not survive indefinitely, but rather collapsed to a black hole. From these arguments, the authors inferred that the remnant collapsed as a SMNS, with a mass near the mass-shedding limit (i.e., the maximum mass that can be supported by uniform rotation). Using a quasi-universal relation between the $M_{\rm max}$ and the mass-shedding limit, they concluded $M_{\rm max} \lesssim 2.16 \substack{+0.17 \\-0.15} \Ms$. 

Reference~\cite{Shibata2017} used numerical relativity simulations to similarly infer that the remnant must have been a longer-lived massive neutron star, but surrounded by a torus. In this scenario, the remnant collapses to a black hole within milliseconds to seconds, depending on the EOS. The authors found that strong neutrino irradiation from a relatively long-lived remnant could provide one way to avoid contamination by lanthanide elements along the line of sight, and could explain the inferred low ejecta opacity. While the presence of a long-lived remnant supports a large $M_{\rm max}$, the absence of optical counterparts from the ejecta restricts the mass from being too large. The authors thus concluded that $M_{\rm max} \sim 2.15-2.25 \Ms$.

Finally, Ref.~\cite{Ruiz2018} used GRMHD simulations to argue that the observed GRB can be explained by a HMNS that undergoes delayed collapse. In this scenario, the remnant mass can be supported by differential rotation as a HMNS, but its mass exceeds that which can be supported through uniform rotation. Using a set of quasi-universal relations inferred from GRMHD simulations for the ratio of the threshold mass for prompt collapse of a HMNS to $M_{\rm max}$ and the ratio of the mass for delayed collapse to $M_{\rm max}$, the authors constrained $M_{\rm max} \in (2.16-2.28)~\Ms$.

These estimates of the maximum mass all point towards a common value of $M_{\rm max} \lesssim 2.3~\Ms$, based, in part, on the timescale of the collapse of the remnant to a black hole.
However, some studies have recently argued that the late-time emission from AT2017gfo can be better explained by energy injection from a long-lived neutron star remnant (with a lifetime $\gtrsim$ days) \cite{Ai2018,Geng2018,Li2018,Yu2018}. This scenario implies a very stiff EOS, in order to support a maximum mass $M_{\rm max} \gtrsim2.6~\Ms$ \cite{Yu2018}. Thus, depending on the interpretation of the electromagnetic counterpart, the implications for $M_{\rm max}$ may vary significantly. Continued observations of the lightcurve or of the polarization evolution may help resolve these differing interpretations for GW170817 \cite{Geng2018}. 

In future observations, post-merger gravitational waves from the remnant may provide a clear indication of the remnant's fate. While no such emission was detected for GW170817, it may be detectable for similar events when advanced detectors reach design sensitivity or next-generation detectors come online \cite{Abbott2017d}.

\section{Bayesian inferences of EOS parameters from GW170817 }
\label{sec:Bayes}
I have so far focused on using the neutron star radius (via $\widetilde{\Lambda}$) to constrain the pressure near $1-2~\rns$ or using $M_{\rm max}$ to constrain the pressure at high densities.  In this section, I now turn to the third and final approach that I will discuss: using the strain data from GW170817 to constrain the EOS of neutron stars directly, via Bayesian inference schemes.

Before going further, it is useful to note why such inference schemes are necessary. Were one to have complete knowledge of the observable functions $\Lambda(\rho_c)$ and $m(\rho_c)$, parameterized arbitrarily by the central density of each star, this would uniquely determine the EOS and the functions could be mathematically inverted to go from $\Lambda-m$ space to pressure-density space. This is a standard example of the inverse stellar structure problem \cite{Lindblom2014}, but only applies for single stars. Recently, Ref.~\cite{Lindblom2018} showed that the single-star inverse structure problem can be generalized to apply to binary systems, and that the set of observables $(m_1, m_2, \widetilde{\Lambda})$ can also be inverted to recover the EOS. However, the inversion is successful only in the limit of a large number of observations that span the complete $(m_1, m_2, \widetilde{\Lambda})$-parameter space. Bayesian inference schemes offer a way to constrain the EOS in the sparse-data limit, in which there are fewer observations than parameters describing the EOS, by incorporating Bayesian priors and even combining independent types of observations.

In Bayesian parameter estimation for a neutron star merger, the parameters are typically decomposed into those that represent a point-particle waveform, $\theta_{pp}$, and those that depend on the EOS, $\theta_{\rm EOS}$ \cite{Abbott2018}. The latter set of parameters can be represented, for example, by the binary effective tidal deformability and its correction term, $\widetilde{\Lambda}$ and $\delta \widetilde{\Lambda}$, or by the tidal deformabilities of the component stars, $\Lambda_1$ and $\Lambda_2$. Alternatively, $\theta_{\rm EOS}$ can be represented by a parameterized EOS, such as those that use piecewise polytropes or linear segments \cite{Read2009,Ozel2009,Hebeler2010,Steiner2010,Raithel2016}, spectral expansions of the adiabatic index \cite{Lindblom2010}, or constant sound-speed parameterizations \cite{Alford2013}.\footnote{For a recent discussion of the effectiveness of piecewise polytropes compared to a spectral expansion parameterization in gravitational wave inference, see Ref. \cite{Carney2018}. For a comparison with respect to a sound-speed parametrization, see Ref.~\cite{Greif2019}. Both these studies highlight the problems that discontinuities in a piecewise-polytropic EOS can pose.} Whereas $\S$\ref{sec:tidal} focused on inferences that sampled $\widetilde{\Lambda}$ and related the results to EOS constraints, in this section, I will review Bayesian schemes that sample the EOS function directly. 

Prior to the observation of GW170817, the inference techniques for connecting observables to the EOS had been well developed. In one early study, Ref. \cite{Read2009a} used numerical simulations to show that the difference between a point-particle inspiral and the inspiral of two neutron stars with realistic EOS could be detected at an effective distance of 100 Mpc with Advanced LIGO. They found that, at that distance, EOS predicting neutron stars that differ in radius by 1~km could be be distinguished, and the pressure at $1.85\rns$ could be constrained to within $5\times10^{33}$~dyn/cm$^2$ \cite{Read2009a}. With an improved waveform model calibrated to results from numerical simulations, it becomes possible to distinguish EOS predicting radii that differ by $\sim10\%$ at distances of up to 300 Mpc \cite{Read2013}. Further numerical simulations have found that tidally-corrected effective-one-body models can accurately measure  tidal effects in the early inspiral phase as well \cite{Baiotti2010,Baiotti2011,Bernuzzi2012,Damour2012,Hotokezaka2013}.

While many of these studies used Fisher matrices to determine detectability of tidal effects on the waveform \cite{Read2009a,Read2013,Flanagan2008,Hinderer2010,Damour2012}, the Fisher matrix method relies on high signal-to-noise, which may be difficult to attain for gravitational wave events, even with multiple detections of neutron star-neutron star mergers \cite{Cokelaer2008,Vallisneri2008}. In the first fully Bayesian parameter estimation of a synthetic gravitational waveform, Ref.~\cite{Del-Pozzo2013} showed that a few tens of detections will provide strong constraints on existing EOS models, by fitting $\theta_{\rm EOS}$ for the tidal deformability. Shortly thereafter, Ref.~\cite{Lackey2015} performed a Bayesian inference on synthetic data that directly sampled the parameters of a piecewise polytropic EOS. They showed that the EOS pressures could be constrained to within a factor of 2 at supranuclear densities, with just one year of observing with a three-detector network, assuming 40 events per year and signal-to-noise $>8$ in each detector. Moreover, they found that most of the constraining power would come from the loudest $\sim$5 events \cite{Lackey2015}, yielding an optimistic prediction for Advanced LIGO operations. In a more recent but similar mock analysis, Ref.~\cite{Abdelsalhin2018} showed that a mere three detections of neutron star mergers, observed by a four-detector network (two LIGO, one Virgo, and one KAGRA detector), would lead to the determination of the central pressures of neutron stars to within 10\%, with potential errors in the pressure at $1.85\rns$ as low as 1\%, for optimal scenarios.

In the LIGO-Virgo Collaboration inference of the neutron star radius from GW170817, Ref.~\cite{Abbott2018} performed the Bayesian parameter estimation in both ways:  by sampling the tidal deformability parameters and, in a separate analysis, by directly sampling the parameters of the EOS function. In the former case,  a set of quasi-universal relations were used to link the tidal deformabilities to neutron star radii, and hence the EOS. For the case in which the EOS function was sampled directly, the authors used a spectral parameterization  \cite{Lindblom2010}, and incorporated astrophysical priors on the neutron star maximum mass to further constrain the parameter space. At the 90\% confidence level, they inferred radii of the two stars to be $R_1 = 10.8\substack{+2.0\\-1.7}$~km and $R_2 = 10.7\substack{+2.1\\-1.5}$~km using the tidal deformability sampling, and $R_1 = 11.9\substack{+1.4\\-1.4}$~km and $R_2 = 11.9\substack{+1.4\\-1.4}$~km for the EOS sampling \cite{Abbott2018}. The latter radii are larger as a result of the extra constraint to support a 1.97~$\Ms$ neutron star, which disfavors very soft EOS. Both sets of radii are consistent with the current range of radii of 9.9-11.2~km, inferred by Ref.~\cite{Ozel2016} from the composite set of X-ray radius measurements, low energy nucleon-nucleon scattering data, and the maximum mass constraint (see $\S$\ref{sec:EOS}). 

Using the spectral EOS sampling, the LIGO-Virgo collaboration also reported direct constraints on the pressure, finding the pressure at twice the nuclear saturation density to be $3.5\substack{+2.7\\-1.7} \times 10^{34}$~dyn/cm$^2$, and at $6\rns$  $9.0\substack{+7.9\\-2.6} \times 10^{35}$~dyn/cm$^{2}$, at the 90\% confidence level \cite{Abbott2018}.

In contrast to these analyses, Ref.~\cite{Landry2018} argued for a non-parametric inference of the EOS. In this scheme, non-parametric priors are constructed using Gaussian processes that have been conditioned on a set of realistic EOS. In a Monte Carlo integration, synthetic EOS are drawn from these non-parametric priors and used to predict the tidal deformabilities for given component masses. These macroscopic properties are then compared with the observed data to calculate the EOS likelihoods. The authors argued that this method will more faithfully recreate complicated EOS, which may have phase transitions or other discontinuities. For GW170817, they found the pressure at twice the nuclear saturation density to be $1.35\substack{+1.8\\-1.2}\times10^{34}$~dyn/cm$^2$, for the case in which the prior was only loosely trained on the realistic EOS, and $4.73\substack{+1.4\\-2.5}\times10^{34}$~dyn/cm$^2$, for the case in which the prior was tightly trained on the sample EOS. For GW170817, this method thus produces results consistent with the LIGO-Virgo parametric analysis \cite{Abbott2018}.

\begin{figure}[ht]
\centering
\includegraphics[width=0.48\textwidth]{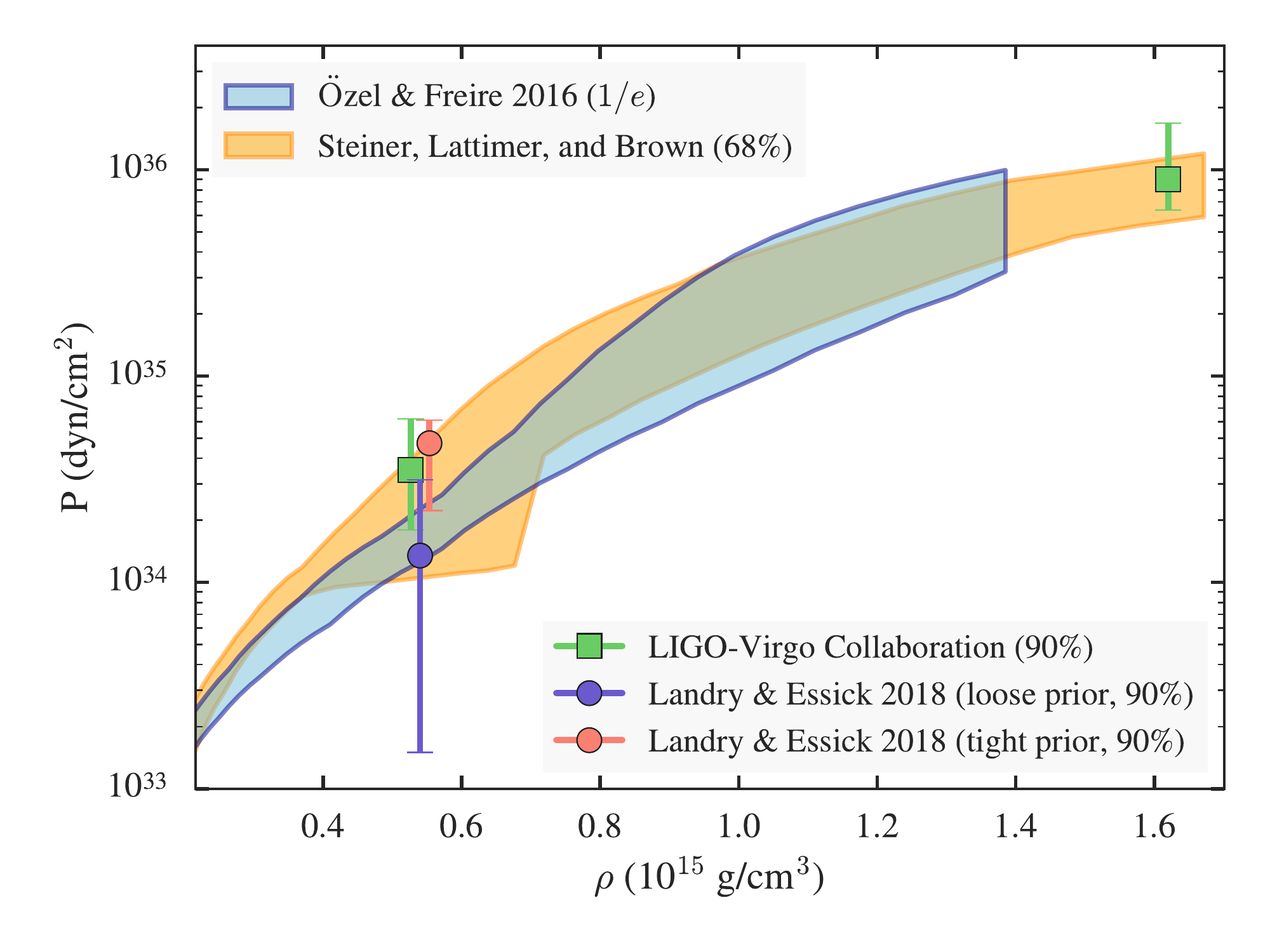}
\caption{\label{fig:EOS} Ensemble of EOS constraints from X-ray radius measurements and GW170817. The orange and blue bands represent the EOS inferred from X-ray measurements of neutron star radii in Refs.~\cite{Steiner2013} and \cite{Ozel2016}, respectively. It should be noted that the blue band spans the range of inferred EOS with posterior likelihoods that fall within $1/e$ of the maximum likelihood, while the orange band represents the 68\% credibility interval. The symbols show the new constraints on the EOS at $2\rns$ and $6\rns$ from analyses of GW170817 \cite{Abbott2018,Landry2018}, with error bars representing the 90\% confidence intervals. While these analyses report their constraints with differing confidences, the results indicate that the constraints from GW170817 are approximately consistent with existing inferences from X-ray measurements. } 
\end{figure}

I summarize all of these results in Fig.~\ref{fig:EOS}. The orange and blue bands represent the EOS inferred from X-ray measurements of neutron star radii in Refs.~\cite{Steiner2013} and \cite{Ozel2016}, respectively (see $\S$\ref{sec:EOS} for details of those analyses). Atop these existing constraints, I show the new constraints on the pressures at $2\rns$ and $6\rns$ from the parametric Bayesian inference of Ref.~\cite{Abbott2018} as well as the non-parametric inference of Ref.~\cite{Landry2018}. While these analyses all report constraints at different confidence levels, it is clear that the two families of observations give approximately consistent results. With additional data-- both of new gravitational wave sources and of new X-ray radii with potentially smaller uncertainties, as are expected from NICER -- the constraints will become even tighter.

\section{Finite-temperature effects}

The EOS constraints discussed in this paper so far have all corresponded to the cold EOS, for which thermal effects are ignored. However, in dynamical phenomena such as neutron star mergers, the temperature may range from this ``cold" regime, to temperatures of up to 10-100~MeV \cite{Oechslin2007}. The additional thermal support provided at these temperatures may affect post-merger gravitational wave frequencies, the lifetime of the remnant, post-collapse accretion disk masses, or other observable features (see, e.g., Ref.~\cite{Baiotti2009}).

If we wish to constrain the cold EOS from observations of these features, we need to be able to disentangle the cold and thermal pressures. A small number of finite-temperature EOS have been formulated to date, which self-consistently calculate thermal effects. Among these are the LS compressible liquid drop model with a Skyrme nuclear force \cite{Lattimer1991}, the Shen relativistic mean field (RMF) thoery model with a Thomas-Fermi approximation \cite{Shen1998a}, and the HS statistical model that has since been applied to $\sim$10 new RMF models and/or mass tables \cite{Hempel2010}. For a recent review on finite-temperature EOS, see Ref.~\cite{Lattimer2016}. However, compared to the diversity of cold EOS models described in $\S$\ref{sec:EOS}, the range of physics probed by finite-temperature EOS is relatively limited.

In order to span a wider range of possible physics, many authors have employed so-called ``hybrid" EOS, in which the thermal pressure of an ideal fluid is added to a cold EOS to account for heating during the merger \cite{Janka1993}. This approach is computationally simple to implement and allows the huge diversity of parametric EOS to be used in representing the cold part of the EOS. However, by representing the thermal component as an ideal fluid, hybrid EOS neglect the significant effects of degeneracy, which become especially important at high densities (for the effects of degeneracy on the thermal pressure, see, e.g., Refs.~\cite{Constantinou2015,Constantinou2015a,Constantinou2017a}). 

Simplifying the thermal effects in this way has important consequences for the inferred observables of mergers. For example, Ref.~\cite{Bauswein2010} showed that using hybrid EOS in numerical simulations of a neutron star-neutron star merger can result in post-merger gravitational wave frequencies that differ by 50-250~Hz from those predicted by more realistic EOS. The lifetime of the hypermassive remnant and post-accretion disk mass following the collapse can also differ significantly.

If progress is to be made in understanding observables from mergers, it is imperative to develop improved techniques for modeling thermal effects. Ideally, such models will allow a wide range of underlying physics in the cold EOS, and will also go beyond the ideal-fluid approximation of hybrid EOS. A first step towards this goal has recently been undertaken in Ref.~\cite{Raithel2019}, in which the authors create a thermal extension that can be added to any cold EOS. They find that degenerate thermal effects can be reasonably approximated with two additional parameters beyond the ideal-fluid model, with errors of $\lesssim30\%$ compared to more realistic calculations. While these results are promising, the full impact of these improved models on merger observables still needs to be explored.

\section{Conclusions}
The discovery of GW170817 has opened up a new era for constraining the neutron star EOS. In particular, the measurement of the tidal deformability of the binary system has unlocked new constraints that map directly to stellar parameters. In this paper, I have reviewed the surprising correlation between $\widetilde{\Lambda}$ and $R$. Using the re-analysis of GW170187 by the LIGO-Virgo collaboration \cite{Abbott2019}, I have shown that the implied radius of the primary neutron star is $9.8 < R < 13.2$~km, for $q=1$, which is consistent with measurements of the radii from X-ray observations \cite{Steiner2013,Ozel2016}. While the $\widetilde{\Lambda}-R$ relationship is a promising tool for hadronic EOS, it seems to break down in the event of a first-order phase transition. GW170817 also inspired new upper limits on the maximum mass of neutron stars, with several independent studies pointing towards $M_{\rm max} \lesssim2.3~\Ms$ and the most restrictive constraint finding, at the 90\% confidence level, $M_{\rm max} \lesssim2.17~\Ms$ \cite{Margalit2017}, under the assumption that the remnant collapses to a neutron star within a few seconds post-merger. These analyses provide more direct observational constraints on the maximum $M_{\rm max}$ than have previously been possible, and can be used to place corresponding upper limits on the EOS pressure at high densities. Additionally, Bayesian parameter estimates allow for direct constraints on EOS, using both parametric and non-parametric methods.  Finally, I reviewed the importance of improving finite-temperature EOS models, in order to be able to disentangle thermal effects from the cold EOS in interpreting future observations of neutron star mergers. 

\acknowledgement{I gratefully acknowledge useful conversations with and feedback from Feryal \"Ozel and Dimitrios Psaltis. This work is supported by NSF Graduate Research Fellowship Program Grant DGE-1143953 and support from NASA grant NNX16AC56G.}


\end{document}